\documentclass[%
 reprint,
 amsmath,amssymb,
 aps,nofootinbib, longbibliography  
]{revtex4-1}
\usepackage{bm}% bold math
\PassOptionsToPackage{linktocpage}{hyperref}
\usepackage[hyperindex,breaklinks]{hyperref}
\usepackage{enumitem}
\usepackage{slashed}
\usepackage{comment}
\usepackage{float}

\renewcommand{\theta}{\vartheta}

\usepackage{soul}

\usepackage{feynmp}
\usepackage{tikz-feynman}
\tikzfeynmanset{compat=1.1.0}

\usepackage{array}
\usepackage{mathtools}
\usepackage{xcolor}

\usepackage{etoolbox}
\smallskipamount

\definecolor{myblue}{rgb}{0.0, 0.0, 0.55}

\begin{document} 
%%%%%%%%%%%%%%%%%%%%%%%%%%%%%%%%%%%%%%%%%%%%%%%%%%%%%%%%
%\title{\textcolor{myblue}{${SO(10)}$, Pati-Salam, Left-Right and all that}}

\title{\textcolor{myblue}{${SO(10)}$ theory on the plateau: the importance of being renormalizable}}
%%%%%%%%%%%%%%%%%%%%%%%%%%%%%%%%%%%%%%%%%%%%%%%%%%%%%%%%

\author{Anca Preda$^{1}$, Goran Senjanovi\'c$^{2,3}$ and Michael Zantedeschi$^{4}$}
\affiliation{%
$^1$
Department of Physics, Lund University, SE-223 62 Lund, Sweden}
\affiliation{
$^2$
Arnold Sommerfeld Center, Ludwig-Maximilians University, Munich, Germany
}%
\affiliation{%
$^3$
International Centre for Theoretical Physics, Trieste, Italy
}%
\affiliation{$^4$ INFN, Sezione di Pisa,
Largo Bruno Pontecorvo 3, I-56127 Pisa, Italy}
\email[]{anca.preda@fysik.lu.se\\ goran.senjanovic@physik.uni-muenchen.de\\ michael.zantedeschi@pi.infn.it}

\begin{abstract}
We revisit a minimal renormalisable $SO(10)$ grand unified theory, with the Higgs representation $45_{\rm H}$, $126_{\rm H}$ and complex $10_{\rm H}$, responsible for the unification, intermediate and the weak scale symmetry breaking, respectively. We perform the study of unification constraints and find that it allows for the Left-Right symmetric scale to be accessible even at the LHC, and the Quark-Lepton unification scale as low as its phenomenological limit around $10^5\,$GeV.
Moreover, one can have neutron - anti neutron oscillations at the level of the present day sensibility in both of the above cases, while in the former case one can have simultaneously neutrinoless double beta decay induced by new light scalar states, reachable today - with both electrons emerging as left-handed, as in the neutrino exchange through its possible Majorana mass. We also discuss a recently raised issue of the fine-tuning of the light Higgs mass and its potential conflict with low intermediate mass scales.
\end{abstract}
\maketitle

\section{Introduction}

Grand unification, from its onset, delivered two fundamental predictions that made it one of the main, if not the main, candidate for the physics beyond the Standard Model: proton decay and the existence of magnetic monopoles. This follows simply from the idea of the unification of the Standard Model gauge forces, which leads to its minimal realisation within the $SU(5)$ gauge group~\cite{Georgi:1974sy}. Although simple extensions can be built - even predictive if based on small representations augmented by higher-dimensional operators~\cite{Ellis:1979fg,Senjanovic:2024uzn,Dorsner:2005fq,Bajc:2006ia} - this theory is tailor-made for massless neutrinos, just like the Standard Model itself. 

If one wanted to unify (a generation of) fermions themselves, on top of interactions, one would have to resort to the $SO(10)$ gauge group~\cite{Fritzsch:1974nn,Georgi:1974my}, which moreover, due to its structure, predicts massive neutrinos, even in its minimal form. This {\it per se}, motivates one to explore this theory as the model of true unification.

There is more to it, though. The proton longevity requires the unification scale to be hopelessly out of direct reach in any foreseeable future, and thus it is natural to look for possible low-energy consequences of grand unification, if they exist at all. For this reason,  
 we have recently revisited a minimal $SO(10)$ grand unified theory with small Higgs representations, augmented by higher-dimensional operators~\cite{Preda:2022izo,Preda:2024upl}. Much to our surprise we found that the phenomenological and theoretical consistency of the theory requires a number of new light scalar states, potentially accessible at today's energies (for a recent review of both minimal non-renormalisable $SU(5)$ and $SO(10)$ theories, see~\cite{Senjanovic:2023jvv}). This is in sharp disagreement with the so-called extended survival principle~\cite{Mohapatra:1982aq} which advocates all scalar states to be as heavy as possible, in accord with residual symmetries. 

The prediction of new light states is deeply tied with the need for higher-dimensional operators that unfortunately introduce a plethora of new couplings. Moreover, this implies that the theory is incomplete which, \textit{per se}, provides a sufficient motivation to turn to a renormalizable version of the $SO(10)$ theory. While this renders the theory complete and more predictive when it comes to fermion masses and 
mixing~\cite{Bajc:2005zf,Babu:2016bmy}, it leads to a large number of new states, which implies the loss of predictivity regarding particle spectra. 

There is still more to it. While it is gratifying to have new light states, it is not the same as having a fully fledged, self-contained theory at energies accessible to present day or near future colliders. In particular, in the context of $SO(10)$, this means that the left-right (LR) symmetric theory~\cite{Pati:1974yy,Mohapatra:1974gc,Senjanovic:1975rk,Senjanovic:1978ev} based on $ SU(2)_{\rm L}\times SU(2)_{\rm R}\times U(1)_{\rm B-L}$ gauge symmetry can be, in principle, reached at the LHC - in fact, the best lower limit comes precisely from the LHC $M_{\rm R}\gtrsim 5 \,\rm TeV$~\cite{ATLAS:2018dcj}. 

If the LR scale were indeed accessible, through the so-called Keung-Senjanovi\'c (KS) process~\cite{Keung:1983uu}, it would lead to lepton number violation consisting of same sign charged di-lepton events and the direct probe of the Majorana nature of heavy right-handed (RH) neutrinos. The former is the high-energy analog of the neutrinoless double beta decay, and the latter feature allows one to untangle the seesaw mechanism~\cite{Nemevsek:2012iq,Senjanovic:2016vxw,Senjanovic:2018xtu,Senjanovic:2019moe}. Moreover,  there is a possible profound connection between the neutrinoless double beta decay and the KS process~\cite{Tello:2010am}. The reviews on the LR theory can be found in~\cite{Senjanovic:2011zz,Tello:2012qda} 
(the reader in need of a more pedagogical {\it expose} is referred to~\cite{Melfo:2021wry}).

The LR symmetric model is naturally partially unified in the context of the Pati-Salam~\cite{Pati:1974yy} gauge group $SU(4)_{\rm C}\times SU(2)_{\rm L}\times SU(2)_{\rm R}$, according to which the leptons are just the fourth color. Although more appealing, unfortunately the quark-lepton unification (QL) scale is too large to be reached by the colliders, since there is 
indirect phenomenological limit $M_{\rm QL}\gtrsim 10^5\,\rm GeV$~\cite{Dolan:2020doe}, still far above the direct experimental reach. Nonetheless,  its physical appeal and the rich phenomenology it offers, makes it worthwhile exploring. 

This is the main focus of our work: we investigate the compatibility of these theories with being phenomenologically and experimentally accessible in the context of $SO(10)$ theory~\cite{Fritzsch:1974nn,Georgi:1974my}. This requires turning to a renormalisable version, since the minimal non-renormalisable  model implies $M_{\rm LR}\simeq M_{\rm QL}\simeq M_{\rm GUT}$~\cite{Preda:2022izo}. 
And indeed the renormalisable version paves the way for low LR and QL scale, albeit at the expense of a proliferation of scalar states and $M_{\rm LR}$ and $M_{\rm QL}$ end up being basically free.
However, the important point is that both the LR and QL scales are allowed to be as low as their phenomenological limits, paving the way for a number of potentially accessible new physical processes. This is against the conventional wisdom of a desert all the way to high energies, and the reason is that in the past most studies utilized one or another version of the extended survival principle~\footnote{There was a claim in the past~\cite{Senjanovic:1982ex}, in the context of extended survival principle, that QL scale could be low in the minimal $SO(10)$, but that required large weak mixing angle, today known to be wrong.}. 

Since these physical scales could actually be accessible even in the context of grand unification with its predictions of proton decay and magnetic monopoles, it is natural to ask if there are additional light states, remnants from the full $SO(10)$ particle spectrum. Similarly, one needs to know whether the unification scale can be low enough to allow for an observable nucleon decay in the next generation experiments. More precisely, we are after the smoking-gun consequence of the fact that these low-energy theories are actually embedded in $SO(10)$.

Thus, besides focusing on the LR and QL theories, we also discuss potentially observable baryon and lepton number violating processes, such as proton decay, neutrinoless double beta decay and neutron-anti neutron oscillations. In the process, we comment on hydrogen-anti hydrogen oscillations and the so-called double proton decay: $ p +  p \to e^+ + e^+$, which unfortunately are far from experimental reach independently of the theoretical framework - but more about it below. 

Before we proceed to demonstrate these results, in order to set the stage, and ease the reader's pain, we start first with a discussion of the partial Pati-Salam unification $SU(4)_{\rm C}\times SU(2)_{\rm L}\times SU(2)_{\rm R}$. For the sake of completeness, we analyse both the minimal non-renormalizable and renormalizable versions. 
In the former case the lower bound on the intermediate scale is $M_{\rm PS}\gtrsim 10^{13}\rm GeV$, while 
in the latter case it can be lowered down to $M_{\rm PS}\gtrsim 10^9\rm GeV$. Clearly, both cases are hopelessly out of experimental reach, which makes the $SO(10)$ theory, besides its beauty and rich new physics, more appealing even from this point of view.

An important comment is in order. When this paper was being readied for publication, we became aware of the result~\cite{Jarkovska:2023zwv} that claims that the minimal renormalisable $SO(10)$ theory could be in trouble, due to the impossibility of fine-tuning the SM Higgs boson. If true, this would be a dramatic result, ruling out the theory. We believe that an independent confirmation is called for before the final verdict is pronounced, especially in view of the death penalty, and thus we decided to report our findings. In order to offer perspective of our results, in view of their assertions, in Sec.~\ref{sec:finetuning} we offer some possible remedies in case their analysis was correct.

\section{Pati-Salam theory}
The Pati-Salam model is based on the $G_{\rm PS}=SU(4)_{\rm C}\times SU(2)_{\rm L}\times SU(2)_{\rm R}$ gauge symmetry augmented by the left-right symmetric representations between $SU(2)_{\rm L}$ and $SU(2)_{\rm R}$.

The minimal fermion content of the model requires two left-right symmetric representation 
\begin{equation}
\label{eq:PSfermion}
    f_{L}= (4_{\rm C},2_{\rm L},1_{\rm R}), \quad f_{R}= (4_{\rm C},1_{\rm L},2_{\rm R}),
\end{equation}
per each generation. 

The strictest bound on the unification scale is derived from rare kaon decays, induced by the new gauge bosons of $SU(4)_{\rm C}$, $X_{\rm PS}$. Being leptoquarks, they induce processes of the type
\begin{equation}
\label{eq:mesondec}
    M \rightarrow \overline{l}_{i}+ l_j,
\end{equation}
where $M$ stands for $K,B$ mesons and $l,j$ stands for different charged leptons. In order to be as conservative as possible, one tries to minimize these effects by judicious choice of mixing angles. This was carefully updated in~\cite{Dolan:2020doe}, with the result
\begin{equation}
\label{eq:boundphenops}
    M_{\rm PS}\gtrsim 10^5\,\rm GeV\,.
\end{equation}

The question we wish to address is whether the model in its minimal realization can saturate this bound. We start with the non-renormalisable version, because it possesses simpler Higgs sector than its renormalisable counterpart, and thus enables a less learned reader to follow the rest of the discussion.

\subsection{\textbf{Non renormalizable version }}

As always, in any theory beyond the Standard Model, the choice of Higgs sector depends on a physical picture that one envisions. Here, we follow the road of Majorana neutrino mass with the seesaw scenario, which then requires the following multiplets
\begin{equation}
\label{eq:PShiggs}
    \begin{split}
        \Delta_{\rm L}=(\overline{10}_{\rm C}&,3_{\rm L},1_{\rm R}),\,\quad \Delta_{\rm R}=(10_{\rm C},1_{\rm L},3_{\rm R}),\\
        &\,\,\,\,\Phi_1= (1_{\rm C},2_{\rm L},2_{\rm R})\,.\,
    \end{split}
\end{equation}
The large symmetry breaking $G_{\rm PS}\rightarrow G_{\rm SM}$ is provided by the local minimum~\cite{Mohapatra:1980yp} $\langle\Delta_{\rm L}\rangle=0$, $\langle\Delta_{\rm R}\rangle\simeq M_{\rm PS}$. All the non Standard Model (SM) gauge bosons and right handed neutrino $N$ get a mass proportional to $\langle\Delta_{\rm R}\rangle$. The $\Phi_1$ field then plays the role of the SM Higgs doublet, properly generalised. Since it is the singlet under quark-lepton symmetry, it leads to wrong mass relation $m_e = m_d$, hence the need for higher-dimensional operators~\cite{Ellis:1979fg}. In the absence of such operators, one similarly predicts $m_D = m_u$, where $m_D$ is Dirac neutrino mass matrix.

In this case, the scale is limited by neutrino mass considerations. What happens is that the third generation neutrino Dirac mass is approximately equal to the top-quark mass $m_{ D3}\simeq m_t$, since the higher-dimensional operator contribution is necessarily much smaller. From the seesaw mechanism~\cite{Minkowski:1977sc,Mohapatra:1979ia,Yanagida:1979as,GellMann:1980vs,Glashow:1979nm} one then has, for the third generation neutirno mass, 
\begin{equation}
    m_{\nu}\simeq \frac{m_t^2}{M_{N}}.
\end{equation}
From $m_\nu\lesssim 1\,\rm eV$, one in turn obtains $M_N\gtrsim 10^{13}\,\rm GeV$, which in view of $M_N\lesssim M_{\rm PS}$, leads to the lower limit on the unification scale
\begin{equation}
    \label{eq:mpsnonren}
    M_{\rm PS}^{({\rm nonren})}\gtrsim 10^{13}\,\rm GeV\,,  
\end{equation}
which makes it hopelessly out of direct experimental reach. 

Notice that the Majorana picture is not just appealing, but necessary. Namely, the analogous Dirac version would imply, as stated above, $m_{\nu3} \simeq m_{t}$.

\subsection{\textbf{Renormalizable version }} 

The renormalizable version of this model requires an additional multiplet 
\begin{equation}
\label{eq:ren}
    \Phi_{15}=(15_{\rm C},2_{\rm L},2_{\rm R})\,,
\end{equation}
which for a sufficiently large Yukawa coupling, through the breaking of quark-lepton symmetry, can correct the wrong mass relations. Since now one loses the connection between up quark and neutrino Dirac mass matrices, the limit in \eqref{eq:mpsnonren} does not apply. Therefore, one must perform a study of gauge coupling unification in order to set the limit on the Pati-Salam scale. 

Before going through the nitty-gritty of this program, a comment is called for regarding the choice of the large scale symmetry breaking Higgs sector. Since in this case there is no connection between neutrino Dirac and up quark masses, one could as well choose the Dirac picture. Although maybe less appealing, it is a perfectly accepted physical possibility, requiring, instead of $\Delta_{\rm LR}$ multiplets, the $(4_{\rm C},2_{\rm L},1_{\rm R})$ and $(4_{\rm C},1_{\rm L},2_{\rm R})$. We should stress that the case of the Dirac neutrino mass has some interesting consequences for the right-handed lepton mixing, and even for strong CP violation~\cite{dartagnan}.
Here we opt for the Majorana road, and leave the Dirac picture for future considerations. 

Pati-Salam model implies the unification of color and $B-L$ in $SU(4)_{\rm C}$, and - due to parity - the equality of the $SU(2)$ gauge couplings at $M_{\rm PS}$.  Therefore, from the definition of hypercharge $Y/2 = T_{3 \rm R}+(B-L)/2$, we arrive at 
\begin{equation}
\label{eq:PScondition}
    \frac{1}{\alpha_1} = \frac{3}{5}\frac{1}{\alpha_2} + \frac{2}{5}\frac{1}{\alpha_3},
\end{equation}
where the factors $\sqrt{3/5}$ and $\sqrt{2/3}$ are  needed to properly normalise $Y/2$ and $(B-L)/2$, respectively. Strictly speaking, there is no need to re-normalise the SM hypercharge $Y$ since its corresponding coupling $\alpha'$ does not unify, but we do it anyway for the sake of the $SO(10)$ analysis that follows later. Of course, \eqref{eq:PScondition} is valid at one-loop order; higher-loops effects slightly correct this~\cite{Hall:1980kf,Weinberg:1980wa}. However, this level of precision suffices for the present discussion.

Considering only the Standard Model particles to be light (equivalent to the extended survival principle), the unification scale turns out to be
\begin{equation}
\label{eq:pssm}
     M_{\rm PS}^{\rm surv} = \exp\left\{\frac{\pi}{22}\left(\frac{5}{\alpha_1} -\frac{3}{\alpha_2}-\frac{2}{\alpha_3}\right)\right\}M_{\rm Z}\simeq 5 \cdot 10^{13}\rm GeV,
\end{equation}
where the gauge couplings $\alpha_3 = 8.4^{-1}$, $\alpha_2 = 29.6^{-1}$ and $\alpha_1 = 59^{-1}$ are evaluated at $M_Z$. This is in agreement with above neutrino mass considerations in the minimal non-renormalizable model. However, in this case there are $d=6$ operators which slightly change the normalization of the gauge fields~\cite{Shafi:1983gz} and thus unification constraints in (\ref{eq:PScondition},\ref{eq:pssm}).

A natural question is whether the new Higgs content of the minimal renormalizable model - c.f. \eqref{eq:PShiggs}, \eqref{eq:ren} - can significantly lower the above scale. Taking into account these particle thresholds, condition \eqref{eq:PScondition} leads to
\begin{equation}
\label{eq:mps}
    \begin{split}
     M_{\rm PS}&=\exp\left\{\frac{\pi}{21}\left(\frac{5}{\alpha_1} -\frac{3}{\alpha_2}-\frac{2}{\alpha_3}\right)\right\} M_Z \cdot\\
   &\quad\left( \frac{(1,1,2)^4(1,3,-1)(6,1,\frac{2}{3})(6,1,\frac{4}{3})^9}{(8,2,\pm\frac{1}{2})^8(6,1,\frac{1}{3})(3,2,\pm\frac{1}{6})^4}\cdot\right. \\
   &\qquad\qquad\qquad \left.\cdot \frac{(3,1,\frac{4}{3})^5(3,2,\pm\frac{7}{6})^{12}M_{\rm Z}^2}{(6,3,\frac{1}{3})^{15}(3,3,\frac{1}{3})^{6}}\right)^{\frac{1}{42}}\,,
    \end{split}
\end{equation}
where, as before, gauge couplings are evaluated at $M_Z$. For convenience, particle masses are now denoted  by the quantum numbers of the corresponding sub-multiplet under $SU(3)_{\rm C} \times SU(2)_{\rm L} \times U(1)_{\rm Y}$ (see the Appendix for decomposition of Pati-Salam multiplets into Standard Model quantum numbers). Conjugation of colour representation has been dropped for clarity, and is, anyway, irrelevant for the purpose of the running.

The lowest unification scale compatible with gauge coupling running is obtained by taking the thresholds (or better, particle states) in the numerator to be as light as possible (i.e. around $\rm TeV$), with those in the numerator as heavy as possible (i.e., at $M_{\rm PS}$), leading to the bound
\begin{equation}
\label{eq:psbound}
    M_{\rm PS}^{({\rm ren})}\gtrsim 2 \cdot 10^{9}\,{\rm GeV}\,.
\end{equation}
While much lower than the scale with only the Standard Model particles being light~\eqref{eq:pssm}, it is still hopelessly too large to be relevant for current experiment. However, there is still a possibility of interesting low-energy physics in this theory - light scalars may induce neutrinoless double beta decay, and even dominate over the usual neutrino exchange, even if the outcoming electrons are to be left handed~\cite{Dvali:2023snt}. In this case, though, the unification scale gets increased by roughly two orders of magnitude.

\subsection{\textbf{Topological defects }} 

The high-energy breaking scale of this model admits~\cite{tHooft:1975psz} magnetic monopoles~\cite{tHooft:1974kcl,Polyakov:1974ek} with mass of order $M_{\rm PS}$. Moreover, the breaking of the discrete left-right symmetry leads also to the production of domain walls. 
Both of these defects are produced via the so-called Kibble mechanism~\cite{Kibble:1976sj}, if a cosmological phase transition takes place in the early universe - and one expects at least one such defect per horizon.

The production  of domain walls~\cite{Zeldovich:1974uw} leads to a cosmological disaster, which however can be easily solved by the presence of interactions explicitly violating left-right symmetry~\cite{Vilenkin:1984ib}, even if induced by tiny gravitional effects suppressed by the Planck scale~\cite{Rai:1992xw}. The generated density of magnetic monopoles~\cite{Zeldovich:1978wj,Preskill:1979zi}, on the other hand, seems to be a problem only for a unification scale higher than about $10^{10}\rm GeV$ - assuming the production of a monopole per horizon, their density today would be just barely compatible with the bounds on monopole flux from MACRO experiment~\cite{MACRO:2002iaq}. In this case, in the minimal renormalisable model - unlike in its non-renormalisable counterpart -  magnetic monopoles could as well exist in accord with cosmology. However, ~\cite{Zurek:1985qw} has claimed that the correlation length could be significantly smaller than the horizon, but that depends on the nature of the phase transition, whose study is outside the scope of this paper.  

However, if one uses tiny explicit breaking of the left-right symmetry to get rid of domain walls, in the process the walls could also sweep away the monopoles~\cite{Dvali:1997sa}, 
in which case it would be hard to know what the final monopole density ends up being. 
We may even be left with none of these interesting topological defects. We will tackle potential signatures of this scenario in a future work~\cite{Senjanovic:25inprogress}.

Alternatively, one may appeal to a possibility of symmetry non-restoration at high temperature~\cite{Weinberg:1974hy,
Mohapatra:1979qt,Mohapatra:1979vr,Mohapatra:1979bt}, which can in principle solve both the domain wall~\cite{Dvali:1995cc,Dvali:1996zr} and the monopole~\cite{Dvali:1995cj} problems. Again, the final monopole density is hard to estimate. 
And, if one is willing to go beyond the minimal models, there is always a possibility of inflation~\cite{Guth:1980zm}.

In any case, in view of the impossibility of experimentally reaching $M_{\rm PS}$ in the Pati-Salam theory, one is encouraged to investigate this issue in the true grand unified theory based on $SO(10)$ gauge symmetry~\cite{Fritzsch:1974nn,Georgi:1974my}, to which we now turn. Suffice it to say here that, in spite of the PS scale being large, one can still have neutrinoless double beta decay be induced by possibly light scalars of the theory, dominating over the neutrino exchange~\cite{Dvali:2023snt}.

\subsection{\textbf{Low LR symmetry scale? }}

The reader may wish to know how low could the LR symmetry breaking scale be in Pati-Salam theory. The answer is trivial in the minimal model, since then it is simply one and the same unification scale $M_{\rm LR} = M_{\rm PS}$. However, if one is willing to add an additional Higgs field $(15_{\rm C},1_{\rm L},1_{\rm R})$, one will have $\langle(15_{\rm C},1_{\rm L},1_{\rm R}) \rangle = M_{\rm PS}$, separated from $\langle(10_{\rm C},1_{\rm L},3_{\rm R})\rangle\simeq M_{\rm LR}$, and $M_{\rm LR}$ now depends on the unification constraints. Can $M_{\rm  LR}$ be as low as its experimental limit around $5\,$TeV?

The unification condition from eq.~\eqref{eq:PScondition} translates in this case into
\begin{equation}
    M^{\rm surv}_{\rm PS}=\exp\left\{\frac{2\pi}{27}\left(\frac{5}{\alpha_1} -\frac{3}{\alpha_2}-\frac{2}{\alpha_3}\right)\right\}M_{\rm Z}\left(\frac{M_{\rm Z}}{M_{\rm LR}}\right)^{\frac{17}{27}}\,,
\end{equation}
if one considers the survival principle, where all particles from $(15_{\rm C},1_{\rm L},1_{\rm R})$ and $(10_{\rm C},1_{\rm L},3_{\rm R})$ are taken to be very heavy. If we require $M_{\rm PS}\leq 10^{17}\,{\rm GeV}$ in order to keep us safe and free from gravity, this translates into a bound on the Left-Right symmetric scale, $M_{\rm LR}\geq 10^{8}{\rm GeV}$.

On the other hand, if one abandons the survival principle and includes scalar thresholds into the gauge couplings' RGE evolutions, the unification condition becomes:
\begin{equation}
\begin{split}
     &M_{\rm PS}=\exp\left\{\frac{2\pi}{41}\left(\frac{5}{\alpha_1} -\frac{3}{\alpha_2}-\frac{2}{\alpha_3}\right)\right\}M_{\rm Z}\left(\frac{M_{\rm Z}}{M_{\rm LR}}\right)^{\frac{17}{41}}\\
     &\left[\frac{(\overline{6},1,-4/3)^9(\overline{6},1,2/3)(\overline{3},1,-2/3)(\overline{3},1,4/3)^5}{M_{\rm Z}^{14}(8,1,0)(\overline{6},1,-1/3)}\right]^{\frac{1}{41}}.
\end{split}
\end{equation}
We found solutions, for example with the $(10_{\rm C},1_{\rm L},3_{\rm R})$ multiplet in the $10-100\,{\rm TeV}$ range, that allow for $M_{\rm LR}$ to be as low as $10\,{\rm TeV}$. This is not unique, but it tells us that low $M_{\rm LR}$, accessible even at the LHC, is perfectly possible within the PS theory.

\section{A minimal renormalisable $SO(10)$ model}\label{sec:minimalrenorm}

The minimal $SO(10)$ model contains the three generations of fermion in $16_{\rm F}$ spinor representations. In the PS language
\begin{equation}
16_{\rm F} = f_{\rm L} + f^c_{\rm L}\,.
\end{equation}
Once again the RH neutrinos $N$ are automatic consequence of the group structure. Just as the PS theory, $SO(10)$ is tailor made for neutrino mass through the seesaw mechanism. In what follows, we shall be brief on theoretical ideas behind neutrino mass and new physics, for a more pedagogical {\it expose} see e.g.~\cite{Senjanovic:2011zz}.

A minimal Higgs sector contains the adjoint $45_{\rm H}$ or a symmetric $54_{\rm H}$ representations, needed for the GUT symmetry breaking at the large scale. Both are equally minimal - they lead to different intermediate symmetries - and the choice is purely personal. This is why we say a (and not the) minimal $SO(10)$ model. 
In this work we will focus on the former representation, since our study of Pati-Salam theory indicates strongly that the corresponding scale ought to be large. While we offer some comments regarding the latter, we leave its investigation for future. 

One also needs a $10_{\rm H}$ representation, containing the SM doublet and thus providing the electroweak symmetry breaking. 

More is needed: one must break the $B-L$ gauge symmetry, and this requires either a spinor $16_{\rm H}$ or a five-index antisymmetric representation $126_{\rm H}$. The former requires higher-dimensional operators in order to reproduce the correct fermion mass spectra, while the latter is self-contained at the tree level, which fits the approach taken in this work. 

We have seen in the PS model that the renormalisable theory requires the bi-doublets $\Phi_1$ and $\Phi_{15}$ in order to generate fermion masses. These fields now reside in $10_{\rm H}$ and $126_{\rm H}$ representations, respectively. Moreover, the $\Delta_{\rm L,R}$ scalar fields now belong to the same $126_{\rm H}$ representations. An important realisation was made in~\cite{Babu:1992ia}, where  it was noticed that once $\langle \Delta_{\rm R}\rangle$ is switched on, also $\langle \Phi_{15} \rangle$ gets triggered through the tadpole term. 

Admittedly, we have a baroque Higgs sector, and yet, it is not sufficient. It was shown in~\cite{Bajc:2005zf} that the resulting fermion mass matrices are still not realistic and hence, even more is needed. Here we are then faced with the choice of either adding another $10_{\rm H}$ or, instead, a $120_{\rm H}$ representation~\cite{Babu:2016bmy}. While the former has less fields\footnote{For a numerical fit of the Yukawa sector the interested reader is referred to~\cite{Joshipura:2011nn}. Notice that under the assumption of spontaneous CP violation, the low-energy theory becomes very similar to the original proposal of T. D. Lee~\cite{Lee:1956qn}. Potential relations with p-decay have been addressed in~\cite{Gao:2024xte}.}, the latter possesses less Yukawa couplings. In what follows, we choose the former and in Sec.~\ref{sec:finetuning} comment on the latter. 

We therefore focus on this renormalisable model based on the Higgs content
\begin{equation}
    45_{\rm H}, \quad 126_{\rm H}, \quad \text{  complex }  10_{\rm H}\,.
\end{equation} 
This theory has been re-discussed in recent years precisely from the particle spectrum point of view~\cite{Bertolini:2012im,Bertolini:2012az}.
Here we address the issue of the values of intermediate mass scale and potential new low-energy phenomenology.

The essential point is that, through the quantum loop effects~\cite{Bertolini:2009es}, the original symmetry breaking $\langle 45_{\rm H} \rangle = M_{\rm GUT}$ allows for the following intermediate symmetry gauge groups 
\begin{equation}
    \begin{split}
        &SO(10) \overset{\langle 45_{\rm H} \rangle}{\longrightarrow} SU(2)_{\rm L} \times SU(2)_{\rm R} \times U(1)_{\rm B-L} \doteq \text{LR}\,,\\
        &SO(10) \overset{\langle 45_{\rm H} \rangle}{\longrightarrow} SU(4)_{\rm C} \times SU(2)_{\rm L} \times U(1)_{\rm R} \doteq \text{QL}\,.
    \end{split}
\end{equation}
This is independent of course of the choice of the intermediate scale Higgs, i.e., whether one uses $16_{\rm H}$ or $126_{\rm H}$ at the next stage of symmetry breaking. If one does not reach directly this scale, is there a way to distinguish these two fundamentally different cases? The answer is yes as we will see now when we discuss explicitly the LR and QL cases.

\subsection{\textbf{LR intermediate symmetry}}
\vspace{2pt}

The particle spectrum of the theory can be obtained readily from the PS theory discussed above. In particular, the fermions in \eqref{eq:PSfermion} are just split into quarks and leptons in the usual manner. Regarding the Higgs sector, the only difference from the PS model is that the LH and RH triplets now become color singlets $\Delta_{\rm L,R}^{(1c)}$. These triplets now live in the $126_{\rm H}$ representations, whereas the bi-doublet $\Phi$ is a linear combination of $\Phi_1$ in $10_{\rm H}$ and $\Phi_{15}$ in $126_{\rm H}$. Since the GUT symmetry breaking  $\langle 45_{\rm H}\rangle$ conserves (for a recent study and references, see e.g.,~\cite{Aulakh:2002zr}) charge conjugation symmetry $C$ (which is a finite $SO(10)$ transformation~\cite{Slansky:1980gc,Slansky:1981yr})~\footnote{To this day, many practitioners of the $SO(10)$ theory for some mysterious reason call this a $D$ symmetry.}, what emerges is the conventional LR model, where 
$C$ takes the role of (generalised) parity $P$.

\begin{figure}[t]
    \includegraphics[width=8.4cm]{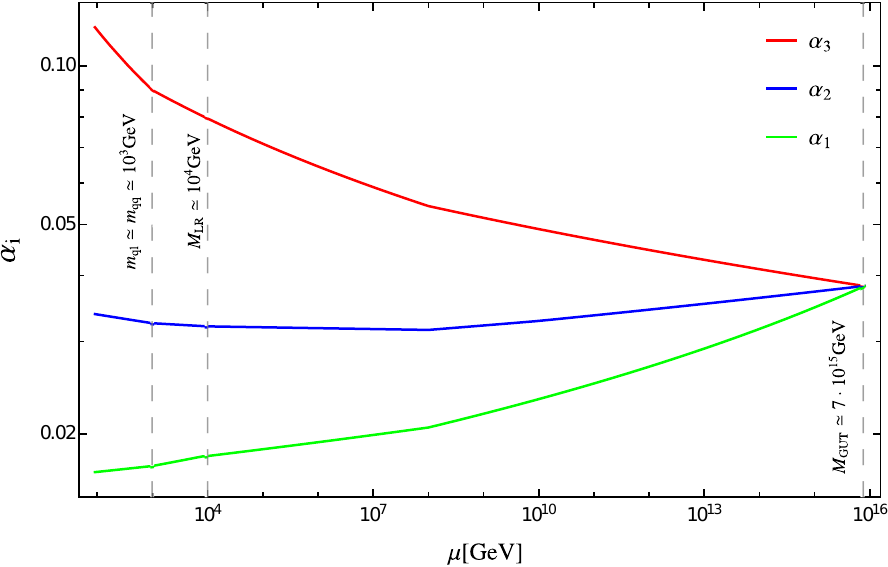}
    \caption{Unification of the SM gauge couplings with an intermediate LR scale $M_{\rm LR}\simeq 10^{4}{\rm GeV}$ and $M_{\rm GUT}\simeq 7\cdot 10^{15}{\rm GeV}$. The fields $\Delta_{ql}$ and $\Delta_{qq}$ lie at $10^{3}{\rm GeV}$, while the second weak doublet in $10_{\rm H}$, as well as the $(1,3,-1)$ and $(1,1,2)$ multiplets have masses around $10^{4}{\rm GeV}$. Additionally, there is also an would-be Goldstone boson $(1,1,1)$, whose mass is precisely at the LR scale. The other fields populating the desert are the $\left(3,2, 1/6\right)$, $\left(\bar{3},2, -1/6\right)$, $\left(3,2, 7/6\right)$, $\left(\bar{3},2, -7/6\right)$ multiplets at $10^{8}{\rm GeV}$ together with the weak triplet in $45_{\rm H}$, whose mass is around $10^{10}{\rm GeV}$. The remaining  scalar states lie at the unification scale.}
    \label{fig:LRrunning}
\end{figure}

Since we are interested here in the low scale LR theory, this implies that the particle masses in the RH multiplets can differ only by roughly $M_{\rm LR}$. Similarly, the same is valid for their LH counterparts.  Specifically, this implies that not only $\Delta_{\rm R}^{(1c)}$ and $\Phi$, but also $\Delta_{\rm L}^{(1c)}$ must live around $M_{\rm LR}$. 

Armed with this, we can proceed to discuss the possible realisation of low lying LR scale consistent with unification constraints. Before that, however, in order to ease the reader's pain, we derive the LR and GUT scale assuming the minimal survival principle. Therefore, the 1-loop renormalization group conditions read
\begin{equation}
    \begin{split}
    &\frac{M_{\rm GUT}}{M_Z}=\exp\left\{\frac{3\pi}{7}\left(\frac{1}{\alpha_2} -\frac{1}{\alpha_3}\right)\right\} \left(\frac{2_{\rm H}\, (1,3,1)^4 }{M_Z^5}\right)^{1/28}\\
    &\,\\
    &\frac{M_{\rm GUT}}{M_Z}=\exp\left\{\frac{30\pi}{127}\left(\frac{1}{\alpha_1} -\frac{1}{\alpha_3}\right)\right\} \\
    &\qquad \qquad \qquad \qquad \left(\frac{2_{\rm H}^5\, (1,3,1)^{18}\,(1,1,3)^{30} M_Z^{79} }{M_{\rm LR}^{132}}\right)^{1/254},
   \end{split}
\end{equation}
where gauge couplings are evaluated at $M_Z$, $2_{\rm H}$ denotes the mass of the second Higgs doublet from $10_{\rm H}$, $(1,3,-1)$ the mass of $\Delta_{\rm L}^{(1c)}$ and $(1,1,3)$ the mass of $\Delta_{\rm R}^{(1c)}$. A similar formula can be derived for the $\alpha_1-\alpha_2$ meeting point but is not needed for our discussion. 

As mentioned above, these states can be at most at the scale $M_{\rm LR}$. In this case, we obtain $M_{\rm LR}\simeq 10^{10}\rm GeV$ and $M_{\rm GUT}\simeq 6\cdot 10^{15}\rm GeV$. 
If, on the other hand, we abandon the survival principle, and let the masses be completely arbitrary, the LR scale can be very different as we are about to show.  

There is, however, an important constraint to keep in mind. Namely, the radiatively corrected spectrum must satisfy a sum rule~\cite{Bertolini:2009es} for the color octet and the weak triplet residing in $45_{\rm H}$: $m_8 + m_3 \gtrsim M_{\rm GUT}$, in an obvious notation. Notice that this is a hard result in the case of our interest, when the LR scale is low and thus cannot affect the sum rule. On the contrary, in the non-renormalisable case this scale ends up lying close to $M_{\rm GUT}$ and thus it manages to effectively eliminate this correlation~\cite{Preda:2022izo}. 

This is rather important since it serves  to distinguish the renormalisable theory from the one based on $16_{\rm H}$ Higgs representation, with the latter possessing a smoking-gun signature of precisely having the color octet and the color triplet (on top of a weak doublet leptoquark) lying at low energies~\cite{Preda:2022izo}. Since the unification scale, where many of the particles lie, cannot be reached directly, it is gratifying to know  that there is a way of knowing whether the theory is based on small or large Higgs representations responsible for the intermediate symmetry breaking.

We illustrate our findings in Fig.~\ref{fig:LRrunning} with a physically interesting  example, where $M_{\rm LR}\simeq 10^{4}\rm GeV$ and moreover, additional light scalars are at TeV energies, potentially accessible at the next generation hadron collider, if not already at the LHC. These are a color sextet, RH triplet and a color triplet, LH triplet, whose role, as we will see below, turns out to be essential for both neutrinoless double beta decay and neutrino - anti neutrino oscillations. 

Last and least, there are additional states along the desert: a color triplet, weak doublet residing in $\Phi_{15}$ with mass around $10^{8}\,\rm GeV$ and a weak triplet from $45_{\rm H}$ with mass around $10^{10}
\,\rm GeV$. 

\subsection{\textbf{QL intermediate symmetry}}
\vspace{2pt}

Although the QL scale cannot be reached directly in near future colliders, it is still of physical interest to investigate whether it can lie close to its lower phenomenological bound. Since now the LR symmetry is broken at the GUT scale, one only has to keep in mind the $SU(4)_{\rm C}$ when discussing the sum rules for scalar masses. In particular, this implies that the multiplet $\Delta_{-1R}^{(\overline{10}C)}$ - a $\overline{10}$ of $SU(4)_{\rm C}$ with minus one $U(1)_{\rm R}$ charge - responsible for the breaking of QL symmetry, cannot be heavier than $M_{\rm QL}$.

In the minimal survival picture scenario, assuming all particle states to be at the possible heaviest stay, we obtain $M_{\rm QL}\simeq 2\cdot 10^{11}\rm GeV$ and $M_{\rm GUT}\simeq 4 \cdot 10^{14}\rm GeV$. These quantities can easily be obtained from the formulas in the Appendix, where the impact of the particle mass thresholds on these scale was derived in general. The reason we obtain an intermediate scale lighter than the LR case previously discussed is due to the impact of the gauge boson $X_{\rm PS}$ on the $\alpha_3$ running, which significantly lowers the meeting point between $\alpha_2$ and $\alpha_3$. Moreover, since $X_{\rm PS}$ contributes also into the $\alpha_1$ gauge coupling running, together with the right handed $W_{\rm LR}$, a slightly higher intermediate scale than the LR case emerges. 

Before commenting on the case where we abandon the survival principle, and vary all masses, it must be mentioned that, as in the case of LR intermediate scale, a similar sum rule emerges~\cite{Bertolini:2009es} for the $SU(4)_{\rm C}$ color fifteen and the weak triplet residing in $45_{\rm H}$: $m_{15} + m_3 \gtrsim M_{\rm GUT}$. As before in the LR case, this clashes with the prediction of the minimal non-renormalisable theory of both the color octet and the weak triplet being accessible at nearby energies~\cite{Preda:2022izo} (see also~\cite{Senjanovic:2022zwy} for some consequences from the latter being light). Once again, we see that, independently of the pattern of symmetry breaking, the theory can differentiate between its $126_{\rm H}$ and $16_{\rm H}$ minimal realizations.

\begin{figure}[t]
    \includegraphics[width=8.4cm]{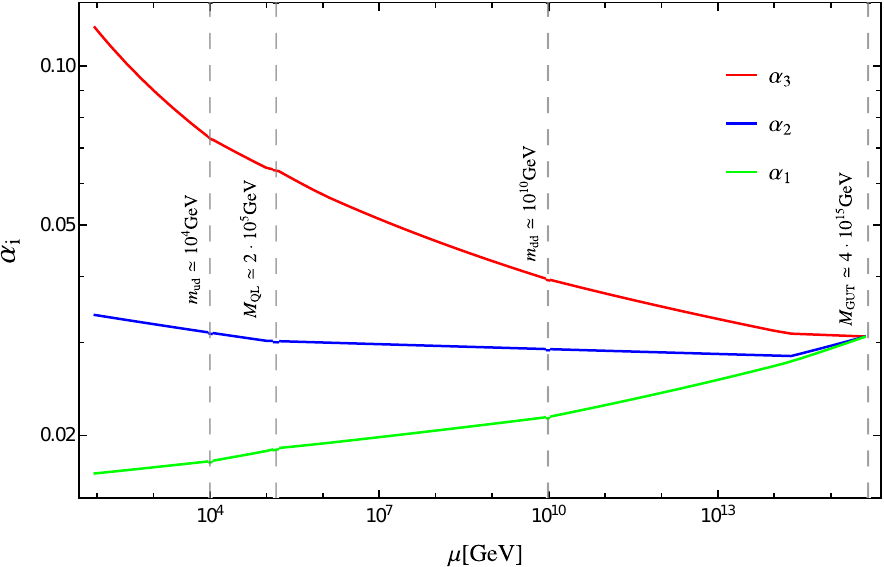}
    \caption{Unification of the SM gauge couplings with an intermediate QL scale $M_{\rm QL}\simeq 2\cdot 10^{5}{\rm GeV}$ and $M_{\rm GUT}\simeq 4\cdot 10^{15}{\rm GeV}$. The field $\Delta_{ud}$, and the scalars $(\bar{3},1,1/3)$ and $(1,1,1)$ corresponding to the same RH multiplet lie at $10^{4}{\rm GeV}$, while $\Delta_{dd}$ and its partners, $(\bar{3},1,4/3)$ and $(1,1,2)$, have masses around $10^{10}{\rm GeV}$. Additionally, $\Delta_{uu}$ lies at $10^{4}{\rm GeV}$ and the would-be Goldstone boson $(\bar{3},1,-2/3)$ lives precisely at the QL scale. The other fields populating the desert are the color octet $(8,1,0)$ and triplets $(3,1,2/3)$, $(\bar{3},1,-2/3)$ from $45_{\rm H}$ with a mass of $10^{6}{\rm GeV}$, the color triplets $(3,1,-1/3)$ from both $10_{\rm H}$ and $126_{\rm H}$, which lie at $10^{14}{\rm GeV}$ and the $(8,2,1/2)$, $(3,2,7/6)$, $(\bar{3},2,-1/6)$, $(1,2,1/2)$ fields at $10^{5}{\rm GeV}$. Finally, the field $\Delta_{\rm L}$ has a mass around $2\cdot 10^{14}{\rm GeV}$ and the remaining  scalar states lie at the unification scale. }
    \label{fig:QLrunning}
\end{figure}

A possible particle spectrum saturating the phenomenological bound \eqref{eq:psbound}, $M_{\rm QL}\simeq 10^{5}\rm GeV$, is shown in Fig.~\ref{fig:QLrunning}. Also in this case, similarly to the LR case, new particle states appear at light energies, in particular arund $10\,\rm TeV$.

\subsection{\textbf{B\&L violating processes}}

More is needed in order to claim that the physics discussed above has its origin in the $SO(10)$ theory. As we know, the essence of grand unification lies in its necessary violation of baryon and lepton numbers (B\&L), that we discuss now.

\paragraph{\bf Nucleon decay}
The prime example of baryon number violation is the $\Delta {\rm B} =1$ nucleon decay. It has two sources, the GUT scale gauge bosons and the heavy scalars, the partners of the SM weak doublets, residing in $10_{\rm H}$ and $126_{\rm H}$ representations. The former is fixed by the GUT scale, while the latter has the freedom of unknown scalar masses. 

Regarding the former. As we have seen, there are a number of examples where the GUT scale is low enough $M_{\rm GUT}\simeq 5\cdot 10^{15}\rm GeV$, providing a potentially visible proton decay. This is a welcome result, but should be taken with a grain of salt. First of all, the proton lifetime scales as the fourth power of the GUT scale, which is extremely sensitive to particle mass thresholds. Moreover, we have other examples with larger GUT scale and in principle, there could be cancellations in nucleon decay amplitudes~\cite{Nandi:1982ew,Dorsner:2004xa,Senjanovic:2024uzn}. A similar situation emerges for the case of the scalar colored triplet, which, in order to be compatible with current bounds on proton lifetime should be heavier than roughly $10^{12}\rm GeV$. Analogously, there are large uncertainties on these masses. 
Thus, by no means our results should be taken as predictions - just a possibility of these decays being reachable in new generation experiments.

\,

What about other interesting B\&L ($\Delta (B - L) =2$) violating processes? Two of them stand out, in a sense that they could be feasible in near future experiments: neutrinoless double beta decay $0\nu2\beta$ and neutron - anti neutron $n-\overline{n}$ oscillations.

\paragraph{\bf Neutron - anti neutron oscillations}

The essential impact of the $126_{\rm H}$ Higgs representation is the breaking of B-L symmetry by 2 units through $\langle \Delta_{\rm R} \rangle$, which results in the Majorana neutrino mass for $N$. In turn, due to quark-lepton symmetry, this implies~\cite{Mohapatra:1980qe} a $\Delta (B-L) = 2$ neutron-anti neutron oscillations. A typical diagram is shown in Fig. 1
For the sake of transparency, it is convenient to introduce the following notation for the color sextet $SU(2)_{\rm R}$ triplet from $\Delta_{\rm R}$ 
\begin{equation}
\label{eq:sextDelta}
\begin{split}
    &\Delta_{uu}=({6}_{\rm C}, \,1_{\rm L},\,Y/2=+4/3),\\ &\Delta_{ud}=( 6_{\rm C}, \,1_{\rm L},\,Y/2=+1/3),\\
    &\Delta_{dd}=( 6_{\rm C}, \,1_{\rm L},\,Y/2=-2/3),
\end{split}
\end{equation}
where in parenthesis we denote the SM quantum numbers.
It should be noted that the $\Delta_{uu}$ is a $SU(4)_{\rm C}$ partner of the Higgs responsible for the QL symmetry breaking, and so it must lie at or below $M_{\rm QL}$, while $\Delta_{ud}$ and $\Delta_{dd}$ should lie at or below $M_{\rm LR}$. We will have to treat separately the two cases in order to be quantitative.

\begin{figure}[t]
\includegraphics[width=6.7cm]{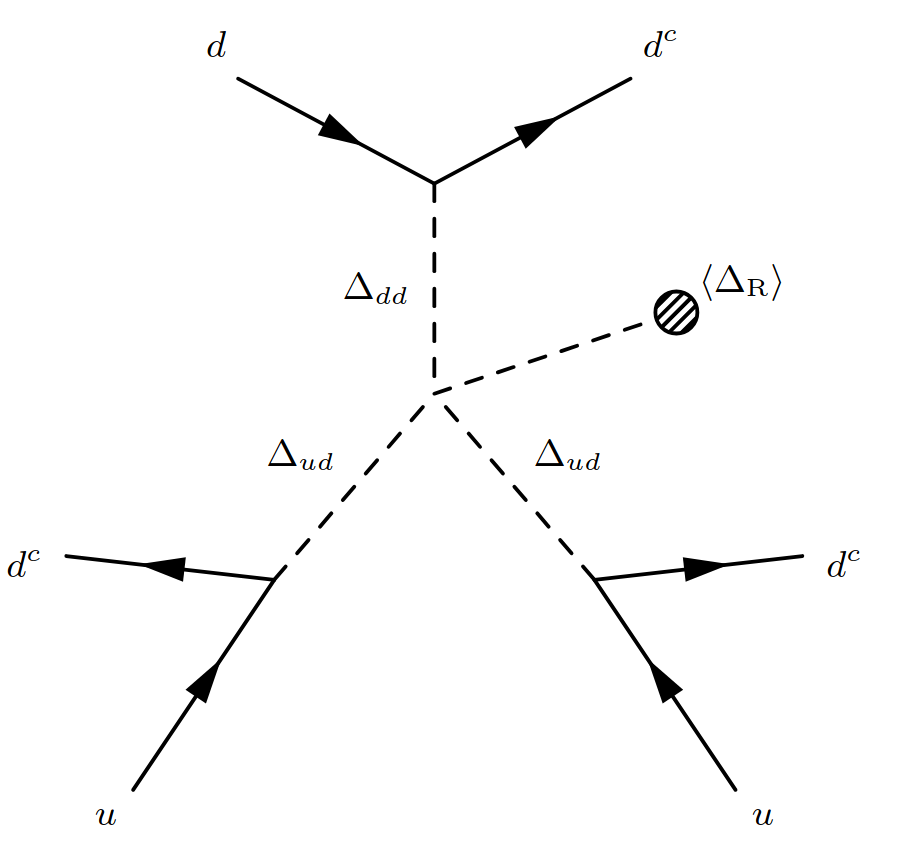}
\label{fig:nnbarosc}
\caption{Example of diagram leading to $n-\overline{n}$ oscillations.}
\end{figure}

The relevant interaction for this process is given by
\begin{equation}\label{eq:nnbarlagr}
   y_{ud}\,\Delta_{ud}\, u_{\rm R}^T\,C\, d_{\rm R} + y_{dd}\,\Delta_{dd}^*\, d_{\rm R}^T\, C \,d_{\rm R} + \alpha\, {\rm{Tr}}\,\Delta_{\rm R}^4\,,
 \end{equation}
where the Yukawa terms originate from the $126_{\rm H}$ interaction with the fermions in $16_{\rm F}$. 

In the following, for the sake of simplicity and illustration we will take these Yukawas to be order one - this will not affect our analysis. Lowering them would imply the relevant mediators to be even lighter, which would put more strain on the consistency of theory. One should keep in mind that these colored scalars masses should be roughly bigger than about TeV. 

The effective operator corresponding to this process - shown in Fig.~1 - has the following form
\begin{equation}
    \frac{1}{\Lambda_{\overline{n}n}^5}d\, d\, u\, d\, d\, u\,,
\end{equation}
which leads to the lifetime (see e.g.~\cite{Babu:2012vc})
\begin{equation}
\label{eq:taunnbar}
    \tau_{n\overline{n}}\simeq \frac{\Lambda_{n\overline{n}}^5}{\Lambda_{\rm QCD}^6}\simeq
    10^{11}{\rm s}\left(\frac{\Lambda_{n\overline{n}}}{10^6\,\rm GeV}\right)^5\,,
\end{equation}
where  
\begin{equation}
\label{eq:lambdannbar}
    \Lambda_{n\overline{n}}^5=\frac{m_{ud}^4\, m_{dd}^2}{\alpha \,\langle\Delta_{\rm R}\rangle},
\end{equation}
and the hadronic
matrix element that takes six quarks to neutrons is $\Lambda_{\rm QCD}^6\simeq 10^{-4}\,\rm GeV^6$ (for the relevant references, see e.g.~\cite{Babu:2012vc}).   
Notice that we could switch $\Delta_{u d}$ for 
$\Delta_{d d}$ and $\Delta_{d d}$ for $\Delta_{u u }$, with similar results. For a lower bound on $n-\overline{n}$ lifetime, roughly of order $10^{8}\,\rm s$, and future prospects, see e.g., Fig.~$33$ in~\cite{Addazi:2020nlz} and references therein. 

\paragraph*{QL case}
From \eqref{eq:taunnbar} and \eqref{eq:lambdannbar}, with $\langle \Delta_{\rm R} \rangle \simeq 10^5\,$GeV and $\alpha \simeq \mathcal{O}(1)$, potentially observable $n-\bar n$ oscillations require 
$m_{ud}^4\, m_{dd}^2 \simeq 10^{35}\, {\rm {GeV}^6}$. It is easy to find solutions with such a constraint, e.g., with $m_{ud} \simeq 10^4\,$GeV and $m_{dd} \simeq 10^{10}\,$GeV. The mass spectrum of the rest of the states is given in Fig.~\ref{fig:QLrunning}.

\paragraph*{LR case}
In this case the situation is less constrained and there are many solutions, including those with  $\Delta_{uu,ud,dd}$ masses as low as $10^3\,$GeV. 
As we will see below, this may be even mandatory if the neutrinoless double beta decay was to be induced by the light scalars, as we now discuss.

\paragraph{\bf Neutrinoless double beta decay}

The canonical contribution to this process comes from the SM $W$-boson exchange and the neutrino Majorana mass which here emerges from the seesaw mechanism, implying the final states electrons to be LH. In the LR model, and thus also in $SO(10)$, there is also an analog $W_{\rm R}$-boson exchange and the Majorana mass for the heavy neutrino $N$~\cite{Mohapatra:1979ia,Mohapatra:1980yp}, in which case the outgoing electrons are RH, and can be in principle distinguished. This is normally ignored, when one assumes the LR scale to be close to the GUT one, but in our case it becomes essential. And, it can be easily distinguished from the neutrino mass induced contribution, since the electron have opposite helicities. 

\begin{figure}[t]
\includegraphics[width=6.3cm]{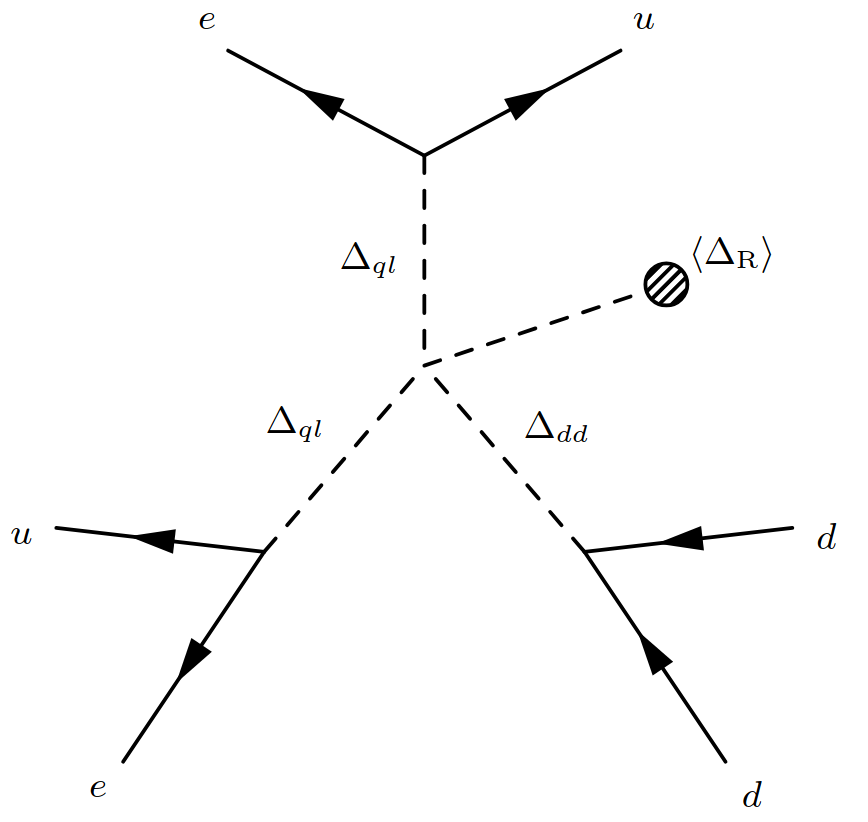}
\label{fig:0ni2b}
\caption{Diagram contributing to $0\nu 2 \beta$.}
\end{figure}
Imagine however that both electrons come out left-handed - would that necessarily imply neutrino Majorana mass as the source of $0\nu2\beta$? This important question was discussed recently in~\cite{Dvali:2023snt}, with the answer in the negative. In this context, the PS model or $SO(10)$ theory are tailor-made to have new physics behind this process. Namely, there are further possible contributions from the colored scalars in $\Delta_{\rm L,R}$ components of $126_{\rm H}$ and depicted in Fig.~2. The relevant couplings obtained from the $126_{\rm H}$ Yukawa interaction with the $16_{\rm F}$ and its scalar potential are given by
\begin{equation}\label{eq:0nu2beta}
  y_{ql}\, q_{\rm L} ^T C \sigma_2 \Delta_{q \ell}  \, \ell_{\rm L}   +  y_{dd}\,\Delta_{d d}^*\, d^{ T}_{\rm R}\,  C\, d_{\rm R}  + \beta\, {\rm{Tr}}(\Delta_{\rm L}^\dagger \Delta_{\rm R})^2\,,
 \end{equation}
where $\Delta_{q \ell} = (\overline{3}_{C},3_{\rm L},Y/2=1/3)$ is a color and weak triplet from $\Delta_{\rm L}$, and $\Delta_{d d}$, a color sextet, weak singlet was defined in \eqref{eq:sextDelta}. In turn, one obtains for the neutrinoless double beta decay effective operator~\cite{Dvali:2023snt} (written in the symbolic form)
\begin{equation}
   \frac{1}{\Lambda^5_{0\nu2\beta}}e^c e^c u^c u^c d\, d,
\end{equation}
where
\begin{equation}
   \Lambda_{0\nu 2\beta}^5= \frac{m_{q \ell}^4\, m_{dd}^2}{\beta \,\langle \Delta_{\rm R}\rangle}\lesssim 10^{18}\,\rm GeV^5\,.
\end{equation}
 The last inequality follows from the latest GERDA results~\cite{GERDA:2020xhi} $\tau_{0\nu2\beta} \geq 10^{26}$yr.
Moreover, both $\Delta_{q \ell}$ and $\Delta_{dd}$ have to be fairly light, even with the Yukawa couplings of order one: $m_{q \ell}\, m_{dd} \lesssim 10^8 \,{\rm GeV}^2$, in order that $0\nu2\beta$ be dominated by these states~\cite{Dvali:2023snt}. 

As manifest from \eqref{eq:0nu2beta} both outgoing electrons are LH, just as in the $W$-neutrino exchange. In order to be quantitative, though, we have to specify the symmetry breaking pattern as in the case of $n-\bar n$ oscillations.

\paragraph*{QL case}
In this case we find no solution with light enough states to allow for the observable $0\nu2\beta$ process. The problem is that the whole $\Delta_{\rm L}$ multiplet then has to lie below the PS scale due to $SU(4)_{\rm C}$ symmetry and this is simply incompatible with the gauge coupling unification with low PS scale.

\paragraph*{LR case}
On the contrary, the LR case allows for both $\Delta_{ql}$ and $\Delta_{qq}$ states to lie as low as TeV, implying a possibility of neutrinoless double decay being induced by these states and not necessarily neutrino Majorana mass - in spite of electrons coming out left-handed. This makes the LR theory tailor-made for this process with a number of competing sources.

There is also a possibility of simultaneously observing $n-\bar n$ oscillations. In fact, with such light di-quarks, the quartic coupling must be kept very small $\alpha \lesssim 10^{-16}$ (c.f., Fig.~1) in order not to violate a present bound on $\tau_{n \bar n}$. Notice that it is completely independent of $\beta$ (c.f., Fig.~2) responsible for $0\nu2\beta$ and so it can be made arbitrarily small. The question is still, whether such a tiny coupling can be theoretically consistent. The answer is yes, as recently discussed~\cite{Dvali:2023snt}. The point is that in the limit $\alpha \to 0$, there is an extra accidental global symmetry, the fermion number, which in the process of symmetry breaking gets traded for a baryon number symmetry. In other words, in the limit $\alpha = 0$, the amplitude for $n-\overline{n}$ oscillations vanishes. Thus we have an example of a technically naturally small quantity, protected to all orders in perturbation theory~\cite{tHooft:1979rat}.

We can have, thus, both of these $\Delta (B-L) =2$ at the same time, without running in contradiction with experiment or the consistency of the theory.

\paragraph{\bf Hydrogen-antihydrogen oscillations}

There are other such processes, as we mentioned in the Introduction: hydrogen - anti hydrogen oscillations and the double proton decay~\cite{Feinberg:1978sd}. They were investigated in~\cite{Arnellos:1982nt,Mohapatra:1982aj,Mohapatra:1982xz}, where it was shown that the mass scale of the respective mediators would have to be no higher than $100$ GeV in order to be potentially observable. 

Notice that hydrogen-antihydrogen oscillations are described by an 8-fermion ($d=12$) effective operator 
\begin{equation}
\label{eq:effHHbar}
    \frac{1}{\Lambda_{\rm \overline HH}^8} d\, u\, u\,e\, d\, u\, u\,e\,,
\end{equation}
thus scaling with the eight power of the effective scale scale $\Lambda_{\overline HH}$. Today the LHC has raised the lower limit on $\Lambda_{\overline HH}$ to about a TeV, indicating that one needs an improvement of experiment by a factor $10^8$, rendering this process hopelessly out of reach

For double proton decay the situation is even grimmer since, being a decay process, its typical timescale goes as the $16^{\rm th}$ power of the analogous effective scale. 

\subsection{\textbf{Topological defects }} 

The GUT scale symmetry breaking leads to the existence of magnetic monopoles, as usual, both in QL and LR cases. According to the Kibble mechanism~\cite{Kibble:1976sj}, which is a conservative lower bound on the production of monopoles, this leads to the infamous over-closure of the universe \cite{Zeldovich:1978wj,Preskill:1979zi} - assuming universe was hot enough for the phase transition to take place. The GUT scale is large enough for the Hubble expansion to dominate over particle rates, so that thermal equilibrium may be lost at such scales~\cite{Elmfors:1993pz} - in other words, there is no reason to claim the monopole problem, as normally done.

 Even if there was an equilibrium, there would a possibility of symmetry non-restoration at these high temperatures, as we mentioned in the case of Pati-Salam theory. The experience with the $SU(5)$ theory tells us that having a large representation, on top of the one responsible for GUT symmetry breaking, helps the GUT symmetry to remain broken above the GUT scale~\cite{Dvali:1995cj}. In our case, $126_{\rm H}$ scalar is tailor made for this job, being much larger than the GUT field $45_{\rm H}$. The precise situation depends, however, on the size of scalar couplings, and it requires a careful analysis, beyond the scope of our work, dedicated to the phenomenological issues.

    The case of LR intermediate symmetry could lead to the existence of cosmologically troublesome domain walls, due to the symmetry breaking of the charge conjugation symmetry at the LR breaking scale. However, since this symmetry is a finite $SO(10)$ global symmetry, the domain walls are not topologically stable. Their fate had been carefully studied in~\cite{Kibble:1982dd,Kibble:1982ae}, however, in an alternative model with $54_{\rm H}$ instead of the $45_{\rm H}$ GUT field. The situation in these cases is rather different due to different intermediate symmetries, and a detailed analysis is called for, along the lines of~\cite{Kibble:1982dd,Kibble:1982ae}. We plan to address this and other aspects of topological defects in the future. \\
    
   One last comment before we close this section. The reader may worry about possible Landau poles on the road to unification, due to the proliferation of light states. She can rest assured, though, that all is well - gauge couplings unify to a perturbative value, and the scalar couplings, being small, stay small all the way to the unification scale. The only danger lies in running above the unification scale, and we comment on it at the end of next section.

\section{\textbf{The issue of weak-scale symmetry breaking}}\label{sec:finetuning}

As we mentioned in the Introduction, in a recent work~\cite{Jarkovska:2023zwv}, the theory was questioned due to a potential problem of too small vev of the weak doublet $\langle 126_{\rm H}\rangle_{(15,2,2)}$ residing in $126_{\rm H}$. This vev ought to be sizable in order to cure the wrong mass relations $m_e = m_d$ and $m_D = m_u$ ($m_D$ is a neutrino Dirac mass matrix) that follow from  $\langle 10_{\rm H} \rangle$.

In order to do that, let us first recapitulate where the problem could stem from. The weak doublet residing in $10_{\rm H}$ must be fine tuned to lie at the weak scale in order to be Higgsed. Now, through the intermediate scale breaking $M_{\rm I}$, the weak doublet in $126_{\rm H}$, with its mass resulting from the high-scale symmetry breaking, obtains a tadpole vev. In other words, in obvious notation
\begin{equation}
\begin{split}
    &\langle 126_{\rm H}\rangle_{(15,2,2)}^{\rm QL} \sim \lambda_{\rm mix}^{\rm QL}\left(\frac{M_{\rm I}}{M_{(15,2,2)}} \right)^2\langle 10_{\rm H}\rangle_{(1,2,2)}\,,\\   
    &   \langle 126_{\rm H}\rangle_{(15,2,2)}^{\rm LR} \sim \lambda_{\rm mix}^{\rm LR}\left(\frac{M_{\rm {GUT}}}{M_{(15,2,2)}} \right)^2\langle 10_{\rm H}\rangle_{(1,2,2)}\,,
\end{split}
\end{equation}
for the QL/LR case of intermediate symmetry respectively, with $\lambda_{\rm mix}^{\rm QL,LR}\lesssim 1$ denoting effective scalar couplings. In the former case, the mixing between the $10_{\rm H}$ and $126_{\rm H}$ doublets goes through $M_{\rm I} =\langle 126_{\rm H}\rangle$, while in the latter it proceeds through $M_{\rm GUT} = \langle 45_{\rm H} \rangle$.
  
 The question is whether $M_{(15,2,2)}$ could end up being too large for the $\langle 126_{\rm H}\rangle_{(15,2,2)}$ vev to matter. In fact,~\cite{Jarkovska:2023zwv} makes a claim of computing the lower limit on the doublet in $126_{\rm H}$, $M_{(15,2,2)}\gtrsim 10^{-1} \, M_{\rm GUT}$. 
If true, for the QL case that would imply that the $M_{\rm I}$ needs to be huge, roughly an order of magnitude below $M_{\rm GUT}$.
The authors of~\cite{Jarkovska:2023zwv}, however, find $M_{\rm I}\lesssim 10^{-3}M_{\rm GUT}$, which would imply too small $\langle 126_{\rm H}\rangle_{(15,2,2)}$ and rule out this case due to the wrong fermion mass relations. 

In the LR case there is no such problem due to the large mixing between the relevant weak doublets, but according to~\cite{Jarkovska:2021jvw} this case suffers from other issues and is ruled out in all of its parameter space. Here we disagree with the claim of~\cite{Jarkovska:2021jvw}, and if we are right, the low LR symmetry scale would work beautifully with the necessary large mixing of the doublets in $10_{\rm H} $ and $126_{\rm H}$.  

We can think of a potential loophole in the analysis of~\cite{Jarkovska:2023zwv}.
Namely, in order to invalidate the theory, the symmetry breaking study requires the Coleman-Weinberg (CW) correction to the tree-level potential, and not as a small perturbation, but on equal footing. The symmetry breaking is in general decided after the one-loop CW contribution is included, since in principle quartic scalar couplings can be small. In other words, the perturbation theory starts at at the one-loop, not tree level. The authors of~\cite{Jarkovska:2023zwv} are of course aware of this and have performed the CW analysis, but did not included the fermions. This however could in principle change their results. The point has to do with the top quark contribution.

Namely, if 
the top quark were to get the mass only from $\langle 10_{\rm H}\rangle$, then we would have for the third generation $m_{D_3}= m_t$, which would require $m_{N_3} \gtrsim 10^{13}\,$GeV, or, in turn, $M_{\rm I}\gtrsim 10^{13}\,$GeV, in contradiction with their claim from the unification constraints. This shows that the top quark has to couple to the doublet in $126_{\rm H}$, which in turn requires including the fermions in the Coleman-Weinberg correction to the tree-level potential. Moreover, their assumption of a relatively small contribution of $126_{\rm H}$ field to the third generation (see below Eq.~$(39)$ in~\cite{Jarkovska:2023zwv}) simply requires a larger $M_{\rm I}$ than claimed by them.
This is yet another reason why we decided to go ahead with publishing our work - simply, we feel that more is needed before this issue is closed for good.

If the indication of the smallness of $\langle 126_{\rm H}\rangle_{(15,2,2)}$ brought by~\cite{Jarkovska:2023zwv} ends up confirmed, it would surely rule out the low intermediate scale physics for the QL case. One should stress that finding particle spectra that allow for $M_{\rm I}$ close to $M_X$ would immediately solve the problem. Indeed, our preliminary studies indicate the possibility of $M_{\rm I} \simeq 10^{14}$\,GeV (with $M_X \simeq 30 M_{\rm I}$), which could do the job, but we leave this issue for a future study. %As we claim in this work, we find the low scale LR intermediate symmetry perfectly consistent, which does not suffer from this issue at all. 
If we are wrong, the fine-tuning problem  
could in principle deal a death blow to the model itself. What would that then imply for the renormalisable $SO(10)$ theory? 

In order to answer this, let us step back, to what the goal actually is - and that is the construction of the minimal renormalisable $SO(10)$ theory. Besides $45_{\rm H}$ and $126_{\rm H}$ representations used for the GUT and intermediate scale symmetry breaking, one needs a real $10_{\rm H}$ field, containing the usual SM Higgs doublet.
 That unfortunately 
fails~\cite{Bajc:2005zf} to reproduce realistic fermion mass matrices, and one natural way out is to add another $10_{\rm H}$, or to complexify it in other words - which brought us to this point. 

The question is whether that is the only way to define this minimal realistic theory. An important question, since physics of spontaneously broken theories depends on the choice of the Higgs scalars, which is not dictated by simple principles. The one theory we have, the SM, does not really indicate which road to take. One lesson could be to take the minimal Higgs representation, but at the end of day we know that it is incomplete, since it predicts vanishing neutrino masses. A simple and predictive remedy is the addition of the complex weak triplet scalar field, which leads to the so-called type II seesaw mechanism~\cite{Magg:1980ut,Lazarides:1980nt,Mohapatra:1980yp}. If one were to opt for this model, the message would be to take all Higgs multiplets that couple to the fermions.

In $SO(10)$ theory, this principle would imply choosing a $120_{\rm H}$ Higgs field~\cite{Babu:2016bmy}, instead of another real $10_{\rm H}$. This is arguably as minimal as the model discussed here, since one should not count the number of particles, but rather the number of representations and use the principle of predictivity when defining minimality - and $120_{\rm H}$ field has actually less Yukawa couplings than $10_{\rm H}$.  

Finally, even if this were to fail, there would be an option of choosing the $54_{\rm H}$ symmetric representation~\cite{Babu:2015bna}, instead of $45_{\rm H}$. This is interesting in itself, since it does not suffer from tree-level breaking problems as $45_{\rm H}$, and the intermediate symmetry, being Pati-Salam, is different - thus the size of $M_{\rm I}$ is expected to change. Moreover, the physics and cosmology of topological defects could change drastically.

There is still a possibility of using the $210_{\rm H}$ as the GUT scale field, which, albeit baroque, offers a different physical picture with charge conjugation symmetry broken at the GUT scale. In other words, the analysis of~\cite{Jarkovska:2023zwv} can serve to choose between minimal renormalisable $SO(10)$ theories, and if even this model were to fail, it would question the whole program of renormalisable $SO(10)$ theory. 

\section{Summary and Outlook}
We have carefully studied a minimal renormalisable $SO(10)$ grand unified theory, based on the following Higgs representations: complex $10_{\rm H}$, 
$45_{\rm H}$ and $126_{\rm H}$, on top of the usual three generations of 
$16_{\rm F}$ fermion representations. The theory can account for all fermion masses, including a naturally small neutrino mass through the seesaw mechanism.  

At the end of the day, we are not completely sure as how to summarise our findings. On one hand, the proliferation of scalar states in $126_{\rm H}$ prevents making any clear prediction regarding the physical mass scales of the theory. One could almost conclude that such a model is not a true theory, not in a sense of being self-contained and predictive.

On the other hand, for the same reason, one has a possibility of populating the 
so-called desert in energies above the weak scale. In particular, the scale of LR symmetry breaking can be accessible even at the LHC and the QL scale of quark-lepton unification can be as low as its phenomenological limit on the order of $10^5 \,{\rm GeV}$. At the same time, there could be additional scalar states lying at low energies.

  In both cases of interest, LR and QL, one could have $n-\bar n$ oscillations at the level of present day experiments. Moreover, in the LR case, one could have also neutrinoless double decay induced by new scalar states, with both electrons coming out LH, just as in the case of Majorana neutrino mass. At the same time, there are realistic prospects for the observable nucleon decay in the forthcoming experiments.

In short, the theory offers a plethora of low energy phenomena, a far cry from the usual desert picture associated with grand unification.
There is a dark cloud on the horizon, though. As the authors of~\cite{Jarkovska:2023zwv} have pointed out, there is an indication that the theory could be ruled out due to the unrealistic weak scale symmetry breaking, combined with the constraints from the unification of gauge couplings. We decided to reserve our judgment before this is verified independently, especially since there is disagreement in some of our results. We believe that the results of~\cite{Jarkovska:2023zwv}, as exciting as they are, are still not conclusive.
We hope, though, that this issue is settled soon.

If the impossibility of achieving realistic weak scale symmetry breaking ends up in a death blow to this version of the minimal renormalisable theory, one will have to resort to an alternative model with a $120_{\rm H}$ Higgs doublet in place of one of the real $10_{\rm H}$ representations, or if even that fails, opt for $54_{\rm H}$ field instead of the adjoint $45_{\rm H}$. This has an appeal in itself, since it leads to the possibility of intermediate symmetry be full Pati-Salam quark-lepton symmetry, discussed in the Section II. This version of the theory does not suffer from tree-level tachyon states present in the $45_{\rm H}$ field, and it possesses a rich spectrum of topological defects. 

In a sense, it could be even nicer if we were wrong - the theory would be far more predictive than imagined originally, or claimed here. A cloud on the horizon may not be that dark, after all. 

If at the end of day, all of these versions of the minimal renormalisable $SO(10)$ turn out not to be realistic, this would provide a boost for a theory with $16_{\rm H}$ in place of $126_{\rm H}$, augmented by higher-dimensional operators~\cite{Preda:2022izo,Preda:2024upl}. Whatever happens, the future of the minimal $SO(10)$ grand unified theory offers excitement.

{\bf Acknowledgments} We are grateful to Borut Bajc for useful discussions and comments. G.S. wishes to acknowledge warm hospitality provided by the friendly staff of the Briig hotel in Split, in whose cafe he found perfect working conditions.

\newpage
\appendix
\section{Decomposition of $SO(10)$ multiplets}
\footnotesize
\renewcommand{\arraystretch}{1.2}
\begin{table}[h!]
\centering
\begin{tabular}{cccc}%{llllllllr}
\hline 
%$10$ %$SO(10)$
 $4_{\rm C}\,2_{\rm L}\,2_{\rm R} $
& $4_{\rm C}\,2_{\rm L}\,1_{\rm R} $
& $3_{\rm C}\,2_{\rm L}\,2_{\rm R}\,1_{BL} $
%& $3_{\rm C}\,2_{\rm L}\,1_{\rm R}\,1_{BL} $
& $3_{\rm C}\,2_{\rm L}\,1_{\rm Y} $
\\
\hline
 $\left({ 1,1,3} \right)$
& $\left({ 1,1},+1 \right)$
& $\left({ 1,1,3},0 \right)$
%& $\left({ 1,1},+1,0 \right)$
& $\left({ 1,1},+1 \right)$
\\
\null
& $\left({ 1,1},0 \right)$
&
%& $\left({ 1,1},0,0 \right)$
& $\left({ 1,1},0 \right)$
\\
\null
& $\left({ 1,1},-1 \right)$
&
%& $\left({ 1,1},-1,0 \right)$
& $\left({ 1,1},-1 \right)$
\\
$\left({ 1,3,1} \right)$
& $\left({ 1,3},0 \right)$
& $\left({ 1,3,1},0 \right)$
%& $\left({ 1,3},0,0 \right)$
& $\left({ 1,3},0 \right)$
\\
$\left({ 6,2,2} \right)$
& $\left({ 6,2},+\frac{1}{2} \right)$
& $\left({ 3,2,2},-\frac{1}{3} \right)$
%& $\left({ 3,2},+\frac{1}{2},-\frac{1}{3} \right)$
& $\left({ 3,2},\frac{1}{6} \right)$
\\
\null
& $\left({ 6,2},-\frac{1}{2} \right)$
&
%& $\left({ 3,2},-\frac{1}{2},-\frac{1}{3} \right)$
& $\left({ 3,2},-\frac{5}{6} \right)$
\\
\null
&
& $\left({ \overline{3},2,2},+\frac{1}{3} \right)$
%& $\left({ \overline{3},2},+\frac{1}{2},+\frac{1}{3} \right)$
& $\left({ \overline{3},2},+\frac{5}{6} \right)$
\\
\null
&
&
%& $\left({ \overline{3},2},-\frac{1}{2},+\frac{1}{3} \right)$
& $\left({ \overline{3},2},-\frac{1}{6} \right)$
\\
$\left({ 15,1,1} \right)$
& $\left({ 15,1},0 \right)$
& $\left({ 1,1,1},0 \right)$
%& $\left({ 1,1},0,0 \right)$
& $\left({ 1,1},0 \right)$
\\
\null
&
& $\left({ 3,1,1},+\frac{2}{3} \right)$
%& $\left({ 3,1},0,+\frac{2}{3} \right)$
& $\left({ 3,1},+\frac{2}{3} \right)$
\\
\null
&
& $\left({ \overline{3},1,1},-\frac{2}{3} \right)$
%& $\left({ \overline{3},1},0,-\frac{2}{3} \right)$
& $\left({ \overline{3},1},-\frac{2}{3} \right)$
\\
\null
&
& $\left({ 8,1,1},0 \right)$
%& $\left({ 8,1},0,0 \right)$
& $\left({ 8,1},0 \right)$
\\
\hline 
\end{tabular}
\caption{Decomposition of $45_{\rm H}$ under PS, QL, LR and SM symmetries with quantum numbers in an obvious notation.}
\label{tab:45decomp}
\end{table}

\renewcommand{\arraystretch}{1.2}
\begin{table}[h!]
\centering
\begin{tabular}{cccc}%{llllllllr}
\hline
%$10$ %$SO(10)$
 $4_{\rm C}\,2_{\rm L}\,2_{\rm R} $
& $4_{\rm C}\,2_{\rm L}\,1_{\rm R} $
& $3_{\rm C}\,2_{\rm L}\,2_{\rm R}\,1_{BL} $
%& $3_{\rm C}\,2_{\rm L}\,1_{\rm R}\,1_{BL} $
& $3_{\rm C}\,2_{\rm L}\,1_{\rm Y} $
\\
\hline
%$126$
$\left({ 6,1,1} \right)$
& $\left({ 6,1,0} \right)$
& $\left({ \overline{3},1,1},+\frac{1}{3} \right)$
%& $\left({ \overline{3},1,0},+\frac{1}{3} \right)$
& $\left({ \overline{3},1},+\frac{1}{3} \right)$
\\
\null
& 
& $\left({ 3,1,1},-\frac{1}{3} \right)$
%& $\left({ 3,1,0},-\frac{1}{3} \right)$
& $\left({ 3,1},-\frac{1}{3} \right)$
\\
\null
$\left({ 10,3,1} \right)$
& $\left({ 10,3,0} \right)$ 
& $\left({ 1,3,1},-1 \right)$
%& $\left({ 1,3,0},-1 \right)$
& $\left({ 1,3},-1 \right)$
\\
\null
& 
& $\left({ 3,3,1},-\tfrac{1}{3} \right)$
%& $\left({ 3,3,0},-\tfrac{1}{3} \right)$
& $\left({ 3,3},-\tfrac{1}{3} \right)$
\\
\null
& 
& $\left({ 6,3,1},+\tfrac{1}{3} \right)$
%& $\left({ 6,3,0},+\tfrac{1}{3} \right)$
& $\left({ 6,3},+\tfrac{1}{3} \right)$
\\
\null
$\left({ \overline{10},1,3} \right)$
& $\left({ \overline{10},1,-1} \right)$ 
& $\left({ 1,1,3},+1 \right)$
%& $\left({ 1,1,-1},+1 \right)$
& $\left({ 1,1},0 \right)$
\\
\null
& $\left({ \overline{10},1,0} \right)$ 
& 
%& $\left({ 1,1,0},+1 \right)$
& $\left({ 1,1},+1 \right)$
\\
\null
& $\left({ \overline{10},1,+1} \right)$ 
& 
%& $\left({ 1,1,+1},+1 \right)$
& $\left({ 1,1},+2 \right)$
\\
\null
& 
& $\left({ \overline{3},1,3},+\tfrac{1}{3} \right)$
%& $\left({ \overline{3},1,-1},+\tfrac{1}{3} \right)$
& $\left({ \overline{3},1},-\tfrac{2}{3} \right)$
\\
\null
& 
& 
%& $\left({ \overline{3},1,0},+\tfrac{1}{3} \right)$
& $\left({ \overline{3},1},+\tfrac{1}{3} \right)$
\\
\null
& 
& 
%& $\left({ \overline{3},1,+1},+\tfrac{1}{3} \right)$
& $\left({ \overline{3},1},+\tfrac{4}{3} \right)$
\\
\null
& 
& $\left({ \overline{6},1,3},-\tfrac{1}{3} \right)$
%& $\left({ \overline{6},1,-1},-\tfrac{1}{3} \right)$
& $\left({ \overline{6},1},-\tfrac{4}{3} \right)$
\\
\null
& 
& 
%& $\left({ \overline{6},1,0},-\tfrac{1}{3} \right)$
& $\left({ \overline{6},1},-\tfrac{1}{3} \right)$
\\
\null
& 
& 
%& $\left({ \overline{6},1,+1},-\tfrac{1}{3} \right)$
& $\left({ \overline{6},1},+\tfrac{2}{3} \right)$
\\
\null
$\left({ 15,2,2} \right)$
& $\left({ 15,2,-\tfrac{1}{2}} \right)$ 
& $\left({ 1,2,2},0 \right)$
%& $\left({ 1,2,-\tfrac{1}{2}},0 \right)$
& $\left({ 1,2},-\tfrac{1}{2} \right)$
\\
\null
& $\left({ 15,2,+\tfrac{1}{2}} \right)$ 
&
%& $\left({ 1,2,+\tfrac{1}{2}},0 \right)$
& $\left({ 1,2},+\tfrac{1}{2} \right)$
\\
\null
&
& $\left({ \overline{3},2,2},-\tfrac{2}{3} \right)$
%& $\left({ \overline{3},2,-\tfrac{1}{2}},-\tfrac{2}{3} \right)$
& $\left({ \overline{3},2},-\tfrac{7}{6} \right)$
\\
\null
&
& 
%& $\left({ \overline{3},2,+\tfrac{1}{2}},-\tfrac{2}{3} \right)$
& $\left({ \overline{3},2},-\tfrac{1}{6} \right)$
\\
\null
&
& $\left({ 3,2,2},+\tfrac{2}{3} \right)$
%& $\left({ 3,2,+\tfrac{1}{2}},+\tfrac{2}{3} \right)$
& $\left({ 3,2},+\tfrac{7}{6} \right)$
\\
\null
&
& 
%& $\left({ 3,2,-\tfrac{1}{2}},+\tfrac{2}{3} \right)$
& $\left({ 3,2},+\tfrac{1}{6} \right)$
\\
\null
& 
& $\left({ 8,2,2},0 \right)$
%& $\left({ 8,2,-\tfrac{1}{2}},0 \right)$
& $\left({ 8,2},-\tfrac{1}{2} \right)$
\\
\null
& 
&
%& $\left({ 8,2,+\tfrac{1}{2}},0 \right)$
& $\left({ 8,2},+\tfrac{1}{2} \right)$
\\
\hline 
\end{tabular}
\caption{Decomposition of $126_{\rm H}$ under PS, QL, LR and SM symmetries.}
\label{tab:126decomp}
\end{table}

\renewcommand{\arraystretch}{1.2}
\begin{table}[h!]
\centering
\begin{tabular}{cccc}%{llllllllr}
\hline 
%$10$ %$SO(10)$
 $4_{\rm C}\,2_{\rm L}\,2_{\rm R} $
& $4_{\rm C}\,2_{\rm L}\,1_{\rm R} $
& $3_{\rm C}\,2_{\rm L}\,2_{\rm R}\,1_{BL} $
%& $3_{\rm C}\,2_{\rm L}\,1_{\rm R}\,1_{BL} $
& $3_{\rm C}\,2_{\rm L}\,1_{\rm Y} $
\\
\hline
 $\left({ 6,1,1} \right)$
& $\left({ 6,1},0 \right)$
& $\left({ 3,1,1},-\frac{2}{3} \right)$
%& $\left({ 1,1},+1,0 \right)$
& $\left({ 3,1},-\frac{1}{3} \right)$
\\
$\left({ 1,2,2} \right)$
&$\left({ 1,2,\frac{1}{2}} \right)$
&$\left({ 1,2,2,0} \right)$
%& $\left({ 1,1},0,0 \right)$
& $\left({ 1,2,\frac{1}{2}} \right)$
\\
\null
& $\left({ 1,2,-\frac{1}{2}} \right)$
&
%& $\left({ 1,1},-1,0 \right)$
& $\left({ 1,2,-\frac{1}{2}} \right)$
\\
\hline 
\end{tabular}
\caption{Decomposition of $10_{\rm H}$ under PS, QL, LR and SM symmetries.}
\label{tab:10decomp}
\end{table}

\newpage
\normalsize
\begin{widetext}
\section{QL breaking pattern}
Below the unification conditions are reported. Fields mass are indicated with quantum numbers under $4_{\rm C} 2_{\rm L} 1_{\rm R}$ for simplicity. Indeed the multiplets can be further splitted by at most $M_{\rm QL}$. In our numerical realizations this was taken into account. 

The meeting point between $\alpha_1,\alpha_2$ is given by
\begin{equation}
\label{eq:12meetingpoint}
\begin{split}
    &\frac{M_{\rm GUT}}{M_Z}= \exp \left\{\frac{10 \pi}{7}\left( \frac{1}{\alpha_1}-\frac{1}{\alpha_2}\right)\right\}\\ 
    &\quad\left[\frac{(10,1,0)^{6}(10,1,1)^{36}(10,1,-1)^{36}(15,1,0)^4(15,2,+1/2)(15,2,-1/2)(6,1,0)^{2}}{(10,3,0)^{82}\, 3_W^{5} \,2_{\rm H} }\right]^{1/21}
    \left(\frac{M_Z}{M_{\rm QL}}\right)^{88/21}\\
    &\quad\simeq  4.6\cdot 10^{44} \left[\frac{(10,1,0)^{6}(10,1,1)^{36}(10,1,-1)^{36}(15,1,0)^4(15,2,+1/2)(15,2,-1/2)(6,1,0)^{2}}{(10,3,0)^{82}\, 3_W^{5} \,2_{\rm H} }\right]^{1/21},
    \end{split}
\end{equation}
where $(6,1,0)$ takes into account both contributions from $126_{\rm H}$ and $10_{\rm H}$ - in particular, the contribution from the latter leads to proton decay and should therefore taken above about $10^{12}\rm GeV$. 

We are interested in the case $M_{\rm QL}\simeq 10^5\,\rm GeV$
Notice that since QL symmetry is broken by $(10,1,-1)$ the multiplet can be at most at that scale. Moreover, one set of weak doublets from $(15,2,\pm 1/2)$ must lie at most at $M_{\rm QL}$ to ensure the lightness of neutrinos. Finally, the Higgs doublet $2_{\rm H}$ comes from $10_{\rm H}$ in which the SM counterpart already reside. Therefore it can also not be heavier than $M_{\rm QL}$.

Taking all the particle thresholds in \eqref{eq:12meetingpoint} at the highest possible mass scale, compatible with the minimal survival principle, we obtain $M_{\rm GUT}\simeq 2\cdot 10^{20}\rm GeV$ for the meeting point between $\alpha_1$ and $\alpha_2$. This value is clearly too high as it is well above $M_{\rm Pl}$. However notice that it is not too high: A small change of particle thresholds can easily bring $M_{\rm GUT}$ below the strong gravity regime scale of about $10^{17}\,{\rm{GeV}}$. This simple fact should somewhat explain the freedom and arbitrariness that allows the model to have, together with a low $M_{\rm QL}$ scale, also other physical processes.

The unification condition between $\alpha_1$ and $\alpha_3$ gives instead
\begin{equation}
\label{eq:13meetingpoint}
    \begin{split}
        \frac{M_{\rm GUT}}{M_Z}&= \exp \left\{\frac{5 \pi}{21}\left( \frac{1}{\alpha_1}-\frac{1}{\alpha_3}\right)\right\}\\
        &\qquad \qquad\left[\frac{(10,1,1)^{14}(10,1,-1)^{14}2_{\rm H} \,  M_Z^5}{(10,3,0)^{18}(10,1,0)^6(15,1,0)^{4}(15,2,2)^2 (6,1,0)^{2}} \right]^{1/84}\left(\frac{M_Z}{M_{\rm QL}}\right)^{11/42} \\
        &\simeq 4.3 \cdot 10^{15}\left[\frac{(10,1,1)^{14}(10,1,-1)^{14}2_{\rm H} \,  M_Z^5}{(10,3,0)^{18}(10,1,0)^6(15,1,0)^{4}(15,2,1/2)(15,2,-1/2) (6,1,0)^{2}} \right]^{1/84}.
    \end{split}
\end{equation}
Upon insertion of the above values at the intermediate scale, the meeting point is $1.3 \cdot 10^{15}\rm GeV$. 

Finally the $\alpha_2-\alpha_3$ meeting point condition reads
\begin{equation}
\label{eq:23meetingpoint}
    \begin{split}
        \frac{M_{\rm GUT}}{M_Z}&= \exp \left\{\frac{2 \pi}{7}\left( \frac{1}{\alpha_2}-\frac{1}{\alpha_3}\right)\right\}\\
        &\left[ \frac{(10,3,0)^{22}3_W^22_{\rm H} M_Z^3}{(10,1,0)^6(10,1,1)^6(10,1,-1)^6(15,1,0)^4(15,2,1/2)(15,2,-1/2)(6,1,0)^2}\right]^{1/42}\left(\frac{M_{\rm QL}}{M_Z}\right)^{11/21}\\
        &\simeq 6.8 \cdot 10^{9}\left[ \frac{(10,3,0)^{22}3_W^22_{\rm H} M_Z^3}{(10,1,0)^6(10,1,1)^6(10,1,-1)^6(15,1,0)^4(15,2,1/2)(15,2,-1/2)(6,1,0)^2}\right]^{1/42}.
    \end{split}
\end{equation}
Once more, taking the relevant multiplets at the intermediate scale $M_{\rm QL}\simeq 10^{5}\,\rm GeV$ gives a meeting point $1.2 \cdot 10^{12}\,\rm GeV$. 

Such a low meeting point is due to the fact that the intermediate scale gauge boson heavily affect the $\alpha_3$ running, therefore lowering the meeting point.
Proton decay bounds requires approximately $M_{\rm GUT}\gtrsim  10^{15}\,\rm GeV$. Therefore, in the running, to ensure a viable unification, tuning of colored particle states was necessary. 
\end{widetext}
\bibliography{biblio}

\end{document}